\documentclass[prd,showpacs]{revtex4}
\usepackage{amsmath,amssymb}
\usepackage{xcolor}

\numberwithin{equation}{section}
\begin{document}
\title{Fermionic Casimir effect at finite temperature in Horava-Lifshitz theories}
\author{Andrea Erdas}
\email{aerdas@loyola.edu}
\affiliation{Department of Physics, Loyola University Maryland, 4501 North Charles Street,
Baltimore, Maryland 21210, USA}
\begin {abstract} 
In this work, I study the finite temperature Casimir effect due to a massless fermion field that violates Lorentz invariance according to the Horava-Lifshitz theory. I investigate a fermion field that obeys MIT bag boundary conditions on a pair of parallel plates. I carry out this study using the generalized zeta function technique that enables me to obtain the Helmholtz free energy and the Casimir pressure when the Casimir plates are in thermal equilibrium with a heat reservoir at finite temperature. I investigate the cases when the parameter associated with the violation of Lorentz invariance is even or odd and the limits of low and high temperature relative to the inverse of plate distance, examining all possible combinations of the above quantities.
In all scenarios studied, I obtain simple and accurate analytic expressions of the free energy and the temperature-dependent Casimir pressure.
\end {abstract}
\maketitle
\section{ Introduction}
\label{1}
The theoretical prediction of the Casimir effect was made 75 years ago \cite{Casimir:1948dh}. 
In Casimir's initial view, quantum field theory effects produced an attraction between two uncharged conducting plates that face each other. 
The first experimental verifications of the effect \cite{Sparnaay:1958wg} came ten years after Casimir published his work and were loosely consistent with the predictions of his theoretical work. 
The Casimir effect has now been confirmed by many experiments of increasing accuracy \cite{Bordag:2001qi,Bordag:2009zz}. 
Casimir's  original paper studied quantum fluctuations of the electromagnetic field in a vacuum and he discovered they produce an attraction of the plates. Later, it was understood that
quantum fluctuations of other fields also produce Casimir forces, which depend strongly on the boundary conditions and shape of the plates \cite{Boyer:1974,Boyer:1968uf}.
Dirichlet, Neumann or mixed boundary conditions have been extensively investigated when studying the Casimir effect due to scalar or vector fields, but cannot be used for fermion fields \cite{Ambjorn:1981xv}. For fermionic fields within the context of the Casimir effect, one must use bag boundary conditions, that have been initially studied to find a solution to confinement.  \cite{Chodos:1974je,Johnson:1975zp} .

The breaking of Lorentz symmetry has been the subject of vigorous theoretical investigations in the last two decades, with proposals of many models that cause a space-time anisotropy  \cite{Horava:2009uw,Ferrari:2010dj,Ulion:2015kjx} within the larger context of seeking a quantum gravity theory. These models break Lorentz invariance at the Planck scale, with repercussions of the space-time anisotropy that are observable at much lower energy through the Casimir effect. Both scalar fields \cite{Cruz:2017kfo,Cruz:2018bqt} and fermion fields \cite{daSilva:2019iwn} that break Lorentz symmetry have been investigated within the context of modifications to the standard Casimir effect. In Ref. \cite{daSilva:2019iwn}, the authors undertake a detailed analysis of the Casimir effect caused by a massless fermion field that violates Lorentz symmetry according to the Horava-Lifshitz model.

Throughout the years, authors have examined thermal effects on the Casimir effect caused by charged scalar fields \cite{Cougo-Pinto:1998fpo,Erdas:2013jga,Erdas:2013dha} and by Majorana fermion fields \cite{Cheng:2010kc,Erdas:2010mz} in Lorentz-symmetric space-time, and thermal corrections to the Casimir effect caused by a charged scalar field that breaks Lorentz invariance \cite{Erdas:2021xvv}. However, no one has studied the finite temperature  Casimir effect caused by a fermion field that breaks Lorentz invariance. In this paper, I will investigate the Casimir effect at finite temperature due to a massless fermion field
that violates Lorentz invariance according to the Horava-Lifshitz model. I will take this fermion field to satisfy MIT bag boundary conditions on two identical parallel plates facing each other.

In Sec. \ref{2} of this article, I briefly present the model of a fermion field that breaks Lorentz invariance according to the theory of Horava-Lifshitz, discuss how this field will satisfy bag boundary conditions on the Casimir plates and then derive the generalized zeta function for the system when it is in contact with a heat reservoir at finite temperature. In Sec. \ref{3}, I use the zeta function regularization to obtain the free energy of this system and the temperature-dependent Casimir pressure it causes on the plates, and examine the limits of low and high temperature. The Horava-Lifshitz theory associates a parameter with the Lorentz symmetry breaking mechanism, and this parameter is a positive integer called critical exponent. In this section, two separate calculations are needed for even and odd values of the critical exponent, and they are presented in two subsections. I summarize and discuss my results in Sec. \ref{4}.
\section{The model and its zeta function}
\label{2}

In this investigation I use $\hbar=c=1$ and study the Casimir effect of a massless fermion field $\psi$ that violates Lorentz symmetry as described by the Horava-Lifshitz model, presented by Horava in Ref. \cite{Horava:2009uw}. The Lorentz symmetry violating Lagrangian for the field $\psi$ \cite{daSilva:2019iwn,Farias:2011aa,DePaola:1999im} is
\begin{equation}
{\cal L}= {\bar \psi}\left[i\gamma^0\partial_t+i^{\xi}\ell^{\xi -1}(\gamma^i\partial_i)^\xi\right]\psi,
\label{L}
\end{equation}
where $\xi$, the critical exponent, is a positive integer that describes the violation of Lorentz symmetry, while the length parameter $\ell$ is introduced to keep the correct dimensionality of $\cal L$. It is obvious that, when $\xi = 1 $, the above Lagrangian is the standard Dirac Lagrangian for a massless fermion which does not violate Lorentz symmetry. Ref. \cite{daSilva:2019iwn} derives the equation of motion of the fermion field $\psi$ from $\cal L$:
\begin{equation}
\left[i\gamma^0\partial_t+i^{\xi}\ell^{\xi -1}(\gamma^i\partial_i)^\xi\right]\psi=0,
\label{DL}
\end{equation}
which represents a "Lorentz violating" Dirac equation for $\xi > 1$, and its authors obtain all the solutions that obey MIT bag boundary conditions on two identical parallel plates of area $L^2$, one located at $z=0$ and the other at 
$z=a$, with $L\gg a$. They show that all real values of the momentum components $k_x$ and $k_y$ are allowed, while only discrete values of $k_z$ are allowed. When $\xi$ is an odd number,  $k_z$ can only take the following values
\begin{equation}
k_z=\left(n+\frac{1}{2}\right)\frac{\pi}{ a},
\label{kz_odd}
\end{equation}
with $n=0,1,2,\cdots$. If $\xi$ is even, $k_z$ can only take the values below
\begin{equation}
k_z={n\pi\over a},
\label{kz_even}
\end{equation}
with $n=1,2,3,\cdots$. 

I will assume the Casimir plates to be in thermal equilibrium with a heat reservoir at temperature $T$ and will calculate thermal effects on their Casimir energy caused by the Lorentz-violating fermion field. For this system the imaginary time formalism of finite temperature field theory is convenient, and it allows only fermionic field configurations satisfying the following anti-periodic condition:
\begin{equation}
\psi(x,y,z,\tau)=-\psi(x,y,z,\tau+\beta)
\label{antiperiodic}
\end{equation}
for any value of the Euclidean time $\tau$, where $\beta = T^{-1}$ is the anti-periodic length in the Euclidean time axis. The Helmholtz free energy of the fermion field is
\begin{equation}
F=-\beta^{-1}\log \,\det \left(D_{\rm E}|{\cal F}_a\right),
\label{F}
\end{equation}
where $D_{\rm E}$ is the square of the "Lorentz violating" Dirac operator in Euclidean time 
\begin{equation}
D_{\rm E}=-\partial^2_\tau-\ell^{2(\xi-1)}\nabla^{2\xi}
\label{D_E}
\end{equation}
and the symbol ${\cal F}_a$ indicates the set of eigenfunctions of $D_{\rm E}$ which satisfy the condition of Eq. (\ref{antiperiodic}), and have discrete values of $k_z$ given by either Eq. (\ref{kz_odd}) or Eq. (\ref{kz_even}). Notice that $\log \,\det \left(D_{\rm E}|{\cal F}_a\right)$ of Eq. (\ref{F}) is related to the partition function $\cal Z$ of the fermion field-Casimir plates system by the following
\begin{equation}
\log {\cal Z}=\log \,\det \left(D_{\rm E}|{\cal F}_a\right),
\label{logZ}
\end{equation}
and, according to the generalized zeta function technique \cite{Hawking:1976ja}, the partition function is related to the zeta function $\zeta(s)$ of the operator $D_{\rm E}$ by the following
\begin{equation}
\log {\cal Z}=-\zeta'(0).
\label{zeta_1}
\end{equation}
To construct this zeta function I need the eigenvalues of the operator $D_{\rm E}$ that satisfy the anti-periodic condition of Eq. (\ref{antiperiodic}). The eigenvalues are listed below
\begin{equation}
\frac{4\pi^2 }{ \beta^2}\left(m+{1\over 2}\right)^2+\ell^{2(\xi-1)}\left(k_\perp^2+k_z^2\right)^\xi,
\label{eigenvalues1}
\end{equation}   
where $m=0,\pm 1, \pm 2, \cdots$, $k_\perp^2=k^2_x+k^2_y$ with $k_x, k_y \in \Re$, and $k_z$ is given by Eq. (\ref{kz_odd}) when $\xi$ is odd and by Eq. (\ref{kz_even}) when $\xi$ is even. With these eigenvalues, I obtain the following zeta function of $D_{\rm E}$
\begin{equation}
\zeta(s)=\mu^{2s}\sum_{m,n}{L^2\over (2\pi)^2}\int d^2k_\perp\left[\frac{4\pi^2 }{ \beta^2}\left(m+{1\over 2}\right)^2+\ell^{2(\xi-1)}\left(k_\perp^2+k_z^2\right)^\xi\right]^{-s},
\label{zeta_2}
\end{equation}   
where, as it is done routinely when applying the zeta function technique, I use the parameter $\mu$ with dimension of mass \cite{Hawking:1976ja} 
 to keep $\zeta(s)$ dimensionless for all values of $s$. Using the following identity, 
 \begin{equation}
z^{-s}=\frac{1}{ \Gamma(s)}\int_0^\infty dt\, t^{s-1}e^{-zt},
\label{z-s}
\end{equation}
 I write the zeta function as
\begin{equation}
\zeta(s)={\mu^{2s}\over \Gamma(s)}\sum_{m,n}{L^2\over (2\pi)^2}\int_0^\infty dt\, t^{s-1}e^{-\frac{4\pi^2 }{\beta^2}\left(m+{1\over 2}\right)^2 t}\int d^2k_\perp e^{-\ell^{2(\xi-1)}\left(k_\perp^2+k_z^2\right)^\xi t}
\label{zeta_3}
\end{equation}
and, once I do the integration over $k_\perp$, obtain
\begin{equation}
\zeta(s)=-{\mu^{2s}\over \Gamma(s)}
{\ell^{(\xi-1+2s)}\over \Gamma(s-\xi/2)}
\sum_{m,n}{L^2\over \sqrt{\pi}}\int_0^\infty dt\, t^{(s-{3\over 2}-{\xi\over 2})}e^{-\frac{4\pi^2 }{\beta^2}\left(m+{1\over 2}\right)^2 t}
e^{-k_z^2 t}.
\label{zeta_4}
\end{equation}
While it is not possible to evaluate this zeta function exactly by analytical means, it is possible to obtain simple analytic expressions in the asymptotic regimes of high temperature and small plate distance (i.e. low temperature).
\section{Free energy and Casimir pressure}
\label{3}

I find the free energy using Eqs. (\ref{F}), (\ref{logZ}), and (\ref{zeta_1}) and write
\begin{equation}
F=\beta^{-1}\zeta'(0).
\label{F2}
\end{equation}
I will now evaluate the zeta function in closed form for the asymptotic cases of small plate distance (low temperature), $a\ll\beta$, and high temperature, $a\gg\beta$. 

In the low temperature approximation, I do a Poisson resummation of the sum over the index $m$ in Eq. (\ref{zeta_4}), 
\begin{equation}
\sum_{m=-\infty}^\infty e^{-\frac{4\pi^2 }{\beta^2}\left(m+{1\over 2}\right)^2 t}={\beta\over\sqrt{\pi t}}\left[{1\over 2}+\sum_{m=1}^\infty(-1)^m e^{-\beta^2{m^2\over {4t}}}\right]
\label{Poisson1}
\end{equation}
and find
\begin{equation}
\zeta(s)=\zeta_0(s)+\zeta_T(s),
\label{zeta_5}
\end{equation}
where
\begin{equation}
\zeta_0(s)=-{\mu^{2s}\over \Gamma(s)}
{\ell^{(\xi-1+2s)}\over \Gamma(s-\xi/2)}{\beta L^2\over 2\pi}
\sum_{n}\int_0^\infty dt\, t^{(s-2-{\xi\over 2})}
e^{-k_z^2 t}
\label{zeta_6}
\end{equation}
is the vacuum part of the zeta function, and 
\begin{equation}
\zeta_T(s)=-{\mu^{2s}\over \Gamma(s)}
{\ell^{(\xi-1+2s)}\over \Gamma(s-\xi/2)}{\beta L^2\over \pi}
\sum_{m,n}(-1)^m \int_0^\infty dt\, t^{(s-2-{\xi\over 2})}e^{-\beta^2{m^2\over {4t}}}
e^{-k_z^2 t}
\label{zeta_7}
\end{equation}
is the temperature correction to the zeta function. I integrate over the variable $t$ in Eq. (\ref{zeta_6}) and find
\begin{equation}
\zeta_0(s)=-{\mu^{2s}\over \Gamma(s)}
{\ell^{(\xi-1+2s)}\over \Gamma(s-\xi/2)}{\beta L^2\over 2\pi}\Gamma(s-\xi/2-1)
\sum_{n}k_z^{(\xi+2-2s)},
\label{zeta0_1}
\end{equation}
while, in Eq. (\ref{zeta_7}), I change the integration variable, $t\rightarrow {m\beta\over 2 k_z} t$, and obtain
\begin{equation}
\zeta_T(s)=-{\mu^{2s}\over \Gamma(s)}
{\ell^{(\xi-1+2s)}\over \Gamma(s-\xi/2)}{\beta L^2\over \pi}
\sum_{m,n}(-1)^m \left({m\beta\over 2 k_z}\right)^{(s-1-\xi/2)} \int_0^\infty dt\, t^{(s-2-{\xi\over 2})}e^{-\beta k_z{m\over {2}}(t+{1\over t})}.
\label{zetaT_1}
\end{equation}
In order to proceed, I need to insert into Eqs. (\ref{zeta0_1}) and (\ref{zetaT_1}) the explicit form of $k_z$, which is different for odd and even values of $\xi$. I will do that in the next two subsections, one dedicated to investigating even values of $\xi$, and one dedicated to odd values of $\xi$. 

When studying the high temperature limit, I Poisson-resumm the summation over the index $n$ in Eq. (\ref{zeta_4}) and find
\begin{equation}
\sum_{n=1}^\infty e^{-\frac{\pi^2 }{a^2}n^2 t}=-{1\over 2}+{a\over\sqrt{\pi t}}\left[{1\over 2}+\sum_{n=1}^\infty e^{-a^2{n^2\over {t}}}\right],
\label{Poisson2}
\end{equation}
for $\xi$ even, and
\begin{equation}
\sum_{n=0}^\infty e^{-\frac{\pi^2 }{a^2}\left(n+{1\over 2}\right)^2 t}={a\over\sqrt{\pi t}}\left[{1\over 2}+\sum_{n=1}^\infty(-1)^n e^{-a^2{n^2\over {t}}}\right],
\label{Poisson3}
\end{equation}
for odd values of $\xi$. I will continue the high temperature calculation in the next two subsections, one for even $\xi$ and one for odd $\xi$.
\subsection{Free energy and Casimir pressure for even $\xi$}
\label{3_1}
I start here by investigating the low temperature limit of the free energy for even values of $\xi$. When $k_z$ takes the form shown in Eq. (\ref{kz_even}), I evaluate the infinite sum of Eq. (\ref{zeta0_1}) in terms of the Riemann zeta function $\zeta_R(z)$
\begin{equation}
\sum_{n=1}^\infty k_z^{(\xi+2-2s)}=\left({\pi\over a}\right)^{(\xi+2-2s)}\zeta_R(2s-2-\xi),
\label{kz1}
\end{equation}
and, using the following
\begin{equation}
A^{2s}
{\Gamma(s-\xi/2-1)\over \Gamma(s) \Gamma(s-\xi/2)}
\zeta_R(2s-2-\xi)\simeq -{\zeta_R(-2-\xi)\over{1+\xi/2}}s+{\cal O}(s^2),
\label{id1}
\end{equation}
valid for $s\ll 1$, where $A$ is a constant independent of $s$, I find
\begin{equation}
\zeta_0(s)\simeq 
{\beta L^2\over 2\pi}{\ell^{(\xi-1)}\over 1+\xi/2}
\left({\pi\over a}\right)^{(\xi+2)}
\zeta_R(-2-\xi)s+{\cal O}(s^2),
\label{zeta0_2}
\end{equation}
for $s\ll 1$. According to Eq. ({\ref{F2}}), the contribution of $\zeta_0$ to the free energy, i.e. the vacuum Casimir energy $F_0$, is
\begin{equation}
F_0=\beta^{-1}\zeta_0'(0),
\label{F3}
\end{equation}
and, since $\zeta_R(-2-\xi)=0$ for any even $\xi$, I find that $F_0=0$ for any even $\xi$, in agreement with Refs. \cite{daSilva:2019iwn,Erdas:2023wzy}.

Next I evaluate $\zeta_T$ using Eq. (\ref{zetaT_1}). In the double sum, only the term with the lowest values of $m$ and $n$, $m=n=1$, contributes significantly, so I write
\begin{equation}
\zeta_T(s)={\mu^{2s}\over \Gamma(s)}
{\ell^{(\xi-1+2s)}\over \Gamma(s-\xi/2)}{\beta L^2\over \pi}
 \left({\beta a\over 2 \pi}\right)^{(s-1-\xi/2)} \int_0^\infty dt\, t^{(s-2-{\xi\over 2})}e^{-{\beta \pi\over {2a}}(t+{1\over t})}.
\label{zetaT_2}
\end{equation}
Using the saddle point method I evaluate the integral and obtain
\begin{equation}
\zeta_T(s)={\mu^{2s}\over \Gamma(s)}
{\ell^{(\xi-1+2s)}\over \Gamma(s-\xi/2)}{2 L^2\over \sqrt{\pi}}
 \left({\beta a\over 2 \pi}\right)^{(s-1/2-\xi/2)} e^{-{\beta \pi\over {a}}}.
\label{zetaT_3}
\end{equation}
Using the following identity, valid for even $\xi$ and for $s\ll 1$
\begin{equation}
{A^{2s}\over \Gamma(s) \Gamma(s-\xi/2)}
\simeq (-1)^{\xi/2}\left({\xi/ 2}\right)! \,\,s^2+{\cal O}(s^3),
\label{id2}
\end{equation}
I find that the temperature correction to the free energy $F_T=\beta^{-1}\zeta_T'(0)$ also vanishes in the low temperature limit for even $\xi$. Therefore, for any even value of the critical exponent, the free energy and the Casimir pressure vanish in the low temperature limit.

I will examine next the high temperature limit of the free energy for even values of $\xi$. I start by inserting Eq. (\ref{Poisson2}) into Eq. (\ref{zeta_4}) and find
\begin{equation}
\zeta(s)=\tilde{\zeta}_T(s)+\tilde{\zeta}_{a,T}(s),
\label{zeta_8}
\end{equation}
where I neglected the $-{1\over 2}$ term of Eq. (\ref{Poisson2}) because it is independent of $a$ and thus will not contribute to the Casimir pressure, and where
\begin{equation}
\tilde{\zeta}_T(s)=-{\mu^{2s}\over \Gamma(s)}
{\ell^{(\xi-1+2s)}\over \Gamma(s-\xi/2)}{L^2a \over 2\pi}
\sum_{m=-\infty}^\infty\int_0^\infty dt\, t^{(s-2-{\xi\over 2})}
e^{-{4\pi^2\over\beta^2}(m+{1\over 2})^2 t}
\label{zeta_9}
\end{equation}
whose $a$-dependence is only through a multiplicative factor of $a$, and 
\begin{equation}
\tilde{\zeta}_{a,T}(s)=-{\mu^{2s}\over \Gamma(s)}
{\ell^{(\xi-1+2s)}\over \Gamma(s-\xi/2)}{ L^2a\over \pi}
\sum_{m,n} \int_0^\infty dt\, t^{(s-2-{\xi\over 2})}
e^{-{4\pi^2\over\beta^2}(m+{1\over 2})^2 t}e^{-a^2{n^2\over {t}}}
\label{zeta_10}
\end{equation}
where the dependence on the plate distance $a$ is more involved. Following steps that are very similar to those used in the low temperature case, I find
\begin{equation}
\tilde{\zeta}_T(s)\simeq 
{ L^2a\over \pi}{\ell^{(\xi-1)}\over 1+\xi/2}
\left({2\pi\over \beta}\right)^{(\xi+2)}
\zeta_R(-2-\xi)s+{\cal O}(s^2),
\label{zeta_11}
\end{equation}
and
\begin{equation}
\tilde{\zeta}_{a,T}(s)=-{\mu^{2s}\over \Gamma(s)}
{\ell^{(\xi-1+2s)}\over \Gamma(s-\xi/2)}{ 2L^2\over \pi}
 \left({\beta a\over \pi}\right)^{(s-1/2-\xi/2)} e^{-{2 \pi a\over {\beta}}}.
\label{zeta_12}
\end{equation}
Once I use $\tilde{\zeta}_T$ and $\tilde{\zeta}_{a,T}$ to calculate the free energy, I find that $F$ vanishes in the high temperature limit because $\zeta_R(-2-\xi)=0$ for even values of $\xi$, and because of the identity of Eq. (\ref{id2}). 
Clearly, the Casimir pressure also vanishes for even values of $\xi$ in the high temperature limit.
\subsection{Free energy and Casimir pressure for odd $\xi$}
\label{3_2}
I begin here by examining the low temperature limit. When $\xi$ is odd, $k_z$ takes the form shown in Eq. (\ref{kz_odd}), and I obtain the following result for the infinite sum of Eq. (\ref{zeta0_1})
\begin{equation}
\sum_{n=0}^\infty k_z^{(\xi+2-2s)}=\left({\pi\over a}\right)^{(\xi+2-2s)}\zeta_H\left(2s-2-\xi, {1\over 2}\right),
\label{kz2}
\end{equation}
where $\zeta_H(z,w)$ is the Hurwitz zeta function defined as
\begin{equation}
\zeta_H\left(z, w\right)=\sum_{n=0}^\infty (n+w)^{-z}.
\label{Hurwitz}
\end{equation}
Next, I use
\begin{equation}
A^{2s}
{\Gamma(s-\xi/2-1)\over \Gamma(s) \Gamma(s-\xi/2)}
\zeta_H(2s-2-\xi,{1\over 2})\simeq -{\zeta_H(-2-\xi, {1\over 2})\over{1+\xi/2}}s+{\cal O}(s^2),
\label{id3}
\end{equation}
valid for small $s$, and find
\begin{equation}
\zeta_0(s)\simeq 
{\beta L^2\over 2\pi}{\ell^{(\xi-1)}\over 1+\xi/2}
\left({\pi\over a}\right)^{(\xi+2)}
\zeta_H(-2-\xi,{1\over 2})s+{\cal O}(s^2),
\label{zeta0_3}
\end{equation}
for $s\ll 1$. Therefore, the vacuum Casimir energy, i.e. the contribution of $\zeta_0$ to the free energy, is
\begin{equation}
F_0={ L^2\over 2\pi}{\ell^{(\xi-1)}\over 1+\xi/2}
\left({\pi\over a}\right)^{(\xi+2)}
\zeta_H(-2-\xi,{1\over 2}),
\label{F4}
\end{equation}
which can be written in a simpler form that contains $\zeta_R$ and not $\zeta_H$
\begin{equation}
F_0=-{ L^2\over \pi}{\ell^{(\xi-1)}\over 2+\xi}
\left({\pi\over a}\right)^{(\xi+2)}
[1-2^{-(\xi+2)}]\zeta_R(-2-\xi).
\label{F5}
\end{equation}
This result fully agrees with Refs. \cite{daSilva:2019iwn,Erdas:2023wzy} for all odd values of $\xi$.

In order to evaluate $\zeta_T$, I start from Eq. (\ref{zetaT_1}) and proceed as I did in the previous subsection. I retain only the term of the double sum with the lowest values of $m$ and $n$ and then use the saddle point method to integrate over $t$, to find
\begin{equation}
\zeta_T(s)={\mu^{2s}\over \Gamma(s)}
{\ell^{(\xi-1+2s)}\over \Gamma(s-\xi/2)}{2 L^2\over \sqrt{\pi}}
 \left({\beta a\over  \pi}\right)^{(s-1/2-\xi/2)} e^{-{\beta \pi\over {2a}}}.
\label{zetaT_4}
\end{equation}
To find the temperature correction to the free energy, $F_T$, I use Eq. (\ref{F2}) and take advantage of the following approximation for $s\ll 1$,
\begin{equation}
{A^{2s}\over \Gamma(s) \Gamma(s-\xi/2)}
\simeq {s\over\Gamma(-\xi/2)}+{\cal O}(s^2),
\label{id4}
\end{equation}
to obtain
\begin{equation}
F_T=
{2 \over \sqrt{\pi}}{L^2\ell^{(\xi-1)}\over \Gamma(-\xi/2)\beta}
 \left({\pi\over \beta a}\right)^{\xi+1\over 2} e^{-{\beta \pi\over {2a}}},
\label{zetaT_5}
\end{equation}
which, using the explicit form of $\Gamma(-\xi/2)$ for odd $\xi$, can be written as
\begin{equation}
F_T=
{2 \over {\pi}}{ L^2 (\xi !!)\ell^{(\xi-1)}\over \beta}
 \left(-{\pi\over 2\beta a}\right)^{\xi+1\over 2} e^{-{\beta \pi\over {2a}}},
\label{zetaT_6}
\end{equation}
where I used the standard notation $\xi !!= 1\cdot 3\cdot 5\cdot\dots\cdot\xi$. This temperature correction, for $\xi =1$, agrees with the result obtained, for example, in Ref. \cite{Erdas:2010mz}, apart from containing an extra factor of two due to the fact that this paper examines Dirac fermions, while the reference deals with Majorana fermions.

The Casimir pressure is obtained immediately using
\begin{equation}
P_C=
-{1\over L^2}{\partial F\over \partial a},
\label{PC_1}
\end{equation}
where $F=F_0+F_T$. I obtain
\begin{equation}
P_C=-{ \ell^{(\xi-1)}\over \pi^2}
\left({\pi\over a}\right)^{(\xi+3)}[1-2^{-(\xi+2)}]\zeta_R(-2-\xi)
-{\ell^{(\xi-1)} (\xi !!)\over a^2}
 \left(-{\pi T\over 2 a}\right)^{\xi+1\over 2} e^{-{ \pi\over {2aT}}},
\label{PC_2}
\end{equation}
which shows the Casimir pressure in the low temperature limit for odd values of the critical exponent. Notice that I wrote the temperature correction piece in terms of the temperature $T$, instead of its inverse $\beta$. Since $\zeta(-3)$, $\zeta(-7)$, $\cdots$ are all positive, while $\zeta(-5)$, $\zeta(-9)$, $\cdots$ are all negative, the pressure is attractive for $\xi=1,5,9,\cdots$ and repulsive for $\xi=3,7,11,\cdots$. Notice also how the temperature correction weakens the vacuum pressure for all values of $\xi$, since the correction and the leading vacuum term always have opposite sign.

When investigating the high temperature limit, I proceed as in the previous subsection and identify two parts of the zeta function. One part, $\tilde{\zeta}_T(s)$, contains $a$ as a simple multiplicative factor and is given by Eq. (\ref{zeta_11}). A second part, $\tilde{\zeta}_{a,T}(s)$, with a more involved dependence on $a$, is given by Eq. (\ref{zeta_12}) without the overall negative sign. With this zeta function, I obtain the high temperature limit of the free energy for odd values of $\xi$
\begin{equation}
F=-{ L^2a\over \pi^2}{\ell^{(\xi-1)}\over 2+\xi}
\left({2\pi\over \beta}\right)^{(\xi+3)}
[1-2^{-(\xi+2)}]\zeta_R(-2-\xi)+{2 \over {\pi}}{L^2(\xi !!)\ell^{(\xi-1)}\over \beta}
 \left(-{\pi\over 2\beta a}\right)^{\xi+1\over 2} e^{-{2 \pi a\over {\beta}}},
\label{F6}
\end{equation}
where the leading term has a linear dependence on the plate distance and represents a uniform energy density, present for all odd values of $\xi$. This term, when $\xi=1$, is the well known  Stefan-Boltzmann term proportional to $T^4$. The high temperature free energy for $\xi = 1$ is in full agreement with the literature, for example Ref. \cite{Erdas:2010mz} (apart from the already mentioned factor of two). 

I take the derivative of the free energy with respect to $a$ and find the high temperature Casimir pressure, which I write in terms of the temperature $T$ rather than its inverse $\beta$
\begin{equation}
P_C={ 1\over \pi^2}{\ell^{(\xi-1)}\over 2+\xi}
\left({2\pi T}\right)^{(\xi+3)}
[1-2^{-(\xi+2)}]\zeta_R(-2-\xi)+{4(\xi !!)\ell^{(\xi-1)}T^2}
 \left(-{\pi T\over 2 a}\right)^{\xi+1\over 2} e^{-{2 \pi aT}},
\label{PC_3}
\end{equation}
where the leading term is repulsive for $\xi=1,5,9,\cdots$ and attractive for $\xi=3,7,11,\cdots$. Notice that the sign of the $a$-dependent correction is always opposite to that of the leading term and thus the correction weakens the pressure for all odd values of $\xi$.
\section{Discussion and conclusions}
\label{4}

In this work, I use the generalized zeta function technique to study the finite temperature Casimir effect of a Lorentz violating massless fermion field that satisfies bag boundary conditions at the plates. 
I obtained an integral expression for the zeta function at finite temperature (\ref{zeta_4}) and used it to find simple analytic expressions of the Helmholtz free energy and Casimir pressure in the case of low temperature, $T \ll a^{-1}$, and high temperature, $T \gg a^{-1}$, for even values of the critical exponent, in Sec. \ref{3_1}, and for odd values of $\xi$, in Sec. \ref{3_2}.

I find that, when the critical exponent is even, the temperature correction $F_T$ to the vacuum Casimir energy is zero for all values of $\xi$, and confirm that, for even values of the critical exponent, the vacuum Casimir energy vanishes as reported in Ref. \cite{daSilva:2019iwn}. I find that the high temperature free energy is zero for all even values of $\xi$. The Casimir pressure is also zero for all even values of $\xi$, and for low and high temperature. Interestingly, for even values of the critical exponent, the vacuum Casimir energy vanishes \cite{daSilva:2019iwn}, the weak magnetic field correction to the Casimir energy and the strong magnetic field Casimir energy vanish \cite{Erdas:2023wzy}, and the low temperature correction to the Casimir energy and the high temperature Helmholtz free energy also vanish, as I show in this work.

I find that, when the critical exponent is odd and temperature is low, the value of the vacuum Casimir energy I report in Eq. (\ref{F5}), agrees with Refs. \cite{daSilva:2019iwn,Erdas:2023wzy} for all odd values of $\xi$, and I obtain its temperature correction, $F_T$, which I report
in Eq. (\ref{zetaT_6}). The Casimir pressure I obtain for low temperature is attractive for $\xi=1,5,9,\cdots$, and repulsive for $\xi=3,7,11,\cdots$, as shown in Eq. (\ref{PC_2}). The temperature correction I obtained decreases the vacuum attractive or repulsive pressure for all odd values of the critical exponent. Below are the free energy and Casimir pressure in the low temperature limit for the first three odd values of $\xi$:
\begin{itemize}
  \item $\xi =1$
  \begin{equation}
F=-{\pi^2\over 360}\left( {7\over 8}\right){L^2\over a^3}-{L^2 T^2\over a}e^{-{\pi\over 2Ta}},
\label{D_1}
\end{equation}
\begin{equation}
P_C=-{\pi^2\over 120 }\left( {7\over 8}\right){1\over a^4} +{\pi \over 2}{T\over a^3}e^{-{\pi\over 2Ta}}.
\label{D_2}
\end{equation}
 \item $\xi=3$
\begin{equation}
F={\pi^{4}\over 1260 }\left(1-2^{-5}\right){L^2 \ell^2\over a^5}+{3\pi\over 2}{ L^2 \ell^2T^3\over  a^2}e^{-{\pi\over 2Ta}},
\label{D_3}
\end{equation}
\begin{equation}
P_C= {\pi^{4}\over 256 }\left(1-2^{-5}\right){\ell^2\over a^6}-{3\pi^2\over 4}{\ell^2 T^2\over a^4}e^{-{\pi\over 2Ta}}.
\label{D_4}
\end{equation}
 \item $\xi=5$
\begin{equation}
F= -{\pi^{6}\over 1680 }\left(1-2^{-7}\right){L^2\ell^4\over a^7}-{15\pi^2\over 4}{ L^2 \ell^4T^4\over  a^3}e^{-{\pi\over 2Ta}},
\label{D_5}
\end{equation}
\begin{equation}
P_C=
 -{\pi^{6}\over 240 }\left(1-2^{-7}\right){\ell^4\over a^8}+{15\pi^3\over 8}{\ell^4 T^3\over a^5}e^{-{\pi\over 2Ta}}.
\label{D_6}
\end{equation}
\end{itemize}

For odd values of the critical exponent in the high temperature limit, the free energy and pressure I obtain are reported in Eqs. (\ref{F6}) and (\ref{PC_3}) respectively.
Below are $F$ and $P_C$  in the high temperature limit for the first three odd values of $\xi$:
\begin{itemize}
  \item $\xi =1$
  \begin{equation}
F=-{2\pi^2\over 45}\left({7\over 8}\right)L^2aT^4-{L^2 T^2\over a}e^{-{2\pi Ta}},
\label{D_7}
\end{equation}
\begin{equation}
P_C={2\pi^2\over 45}\left({7\over 8}\right)T^4-2\pi{ T^3\over a}e^{-{2\pi Ta}}.
\label{D_8}
\end{equation}
 \item $\xi=3$
\begin{equation}
F={16\pi^{4}\over 315 }\left(1-2^{-5}\right){L^2a\ell^2T^6}+{3\pi\over 2}{ L^2 \ell^2T^3\over  a^2}e^{-{2\pi Ta}},
\label{D_9}
\end{equation}
\begin{equation}
P_C=-{16\pi^{4}\over 315 }\left(1-2^{-5}\right){\ell^2T^6}+3\pi^2 {\ell^2T^4\over  a^2}e^{-{2\pi Ta}}.
\label{D_10}
\end{equation}
 \item $\xi=5$
\begin{equation}
F=-{16\pi^{6}\over 105 }\left(1-2^{-7}\right){L^2a\ell^4T^8}-{15\pi^2\over 4}{ L^2 \ell^4T^4\over  a^3}e^{-{2\pi Ta}},
\label{D_11}
\end{equation}
\begin{equation}
P_C={16\pi^{6}\over 105 }\left(1-2^{-7}\right){\ell^4T^8}-{15\pi^3\over 2}{\ell^4T^5\over  a^3}e^{-{2\pi Ta}}.
\label{D_12}
\end{equation}
\end{itemize}
Notice how the numerical factor that determines the leading term of the Casimir pressure increases as $\xi$ increases, from $0.384$ to $4.79$ to $145.35$.

\end{document}